\documentclass{article}

\usepackage{arxiv}

\usepackage{amsmath}
\usepackage[utf8]{inputenc} 
\usepackage[T1]{fontenc}    
\usepackage{hyperref}       
\usepackage{url}            
\usepackage{booktabs}       
\usepackage{amsfonts}       
\usepackage{nicefrac}       
\usepackage{microtype}      
\usepackage{cleveref}       
\usepackage{lipsum}         
\usepackage{graphicx}
\usepackage{doi}
\usepackage{booktabs}
\usepackage{comment}
\usepackage[ruled,linesnumbered]{algorithm2e}
\usepackage{subfigure}
\usepackage{caption}
\usepackage{multirow}

\title{Analyzing Query Optimizer Performance in the Presence and Absence of Cardinality Estimates}

\newif\ifuniqueAffiliation
\uniqueAffiliationtrue

\ifuniqueAffiliation
\author{Asoke Datta, Brian Tsan, Yesdaulet Izenov, Florin Rusu\\
	University of California Merced\\
	\texttt{\{adatta2, btsan, yizenov, frusu\}@ucmerced.edu} \\
}

\renewcommand{\headeright}{Datta et al.}
\renewcommand{\undertitle}{Technical Report}
\renewcommand{\shorttitle}{Query Optimization in the Presence and Absence of Cardinality Estimates}

\hypersetup{
pdftitle={Query Optimization in the Presence and Absence of Cardinality Estimates},
pdfauthor={Asoke Datta, Brian Tsan, Yesdaulet Izenov, Florin Rusu},
pdfkeywords={Query Optimization, Cardinality Estimation}
}

\begin{document}
\maketitle

\begin{abstract}
Most query optimizers rely on cardinality estimates to determine optimal execution plans. While traditional databases such as PostgreSQL, Oracle, and Db2 utilize many types of synopses -- including histograms, samples, and sketches -- recent main-memory databases like DuckDB and Heavy.AI often operate with minimal or no estimates, yet their performance does not necessarily suffer. To the best of our knowledge, no analytical comparison has been conducted between optimizers with and without cardinality estimates to understand their performance characteristics in different settings, such as indexed, non-indexed, and multi-threaded. In this paper, we present a comparative analysis between optimizers that use cardinality estimates and those that do not. We use the Join Order Benchmark (JOB) for our evaluation and true cardinalities as the baseline. Our investigation reveals that cardinality estimates have marginal impact in non-indexed settings. Meanwhile, when indexes are available, inaccurate estimates may lead to sub-optimal physical operators --- even with an optimal join order. Furthermore, the impact of cardinality estimates is less significant in highly-parallel main-memory databases.
\end{abstract}


\section{INTRODUCTION}\label{sec:intro}

Accurate cardinality estimation is vital to query optimization. However, Leis et al.\cite{Leis:QOREALLY:pvldb-2015,Leis:JOB:vldb-2018} observed, "when the database has only primary key indexes, and once nested-loop joins have been disabled and rehashing has been enabled, the performance of most queries is close to the one obtained using the true cardinalities. Given the bad quality of the cardinality estimates, we consider this to be a surprisingly positive result". This outcome arises mainly because the lack of indexes on the fact tables necessitates costly full-table scans, diminishing the performance gap between an optimal and a suboptimal join order. Nonetheless, it remains essential to avoid joining two fact tables, for which even inaccurate cardinalities may suffice \cite{Leis:QOREALLY:pvldb-2015,Leis:JOB:vldb-2018}.

\begin{figure}[htbp]
	\centering
	\includegraphics[width=0.85\textwidth]{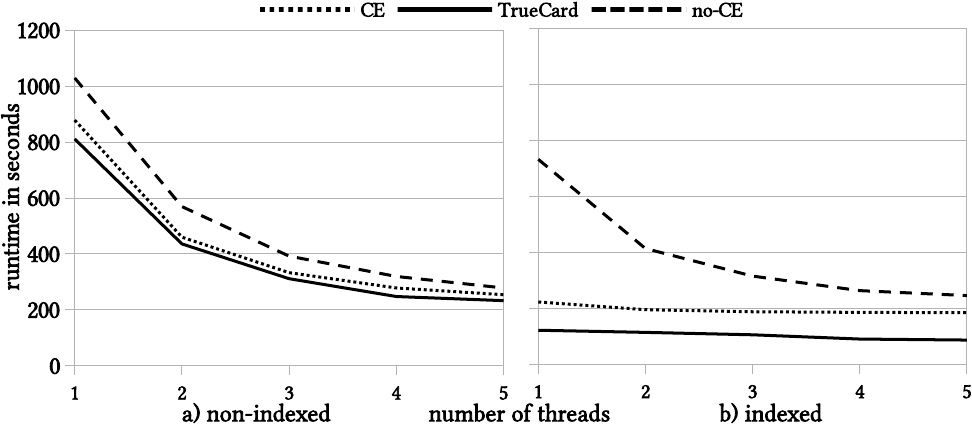}
	\caption{JOB workload runtime across number of threads ranging from 1-5.}
	\label{fig:motivation}
\end{figure}

We find it noteworthy that the gap between optimal and suboptimal query plans diminishes, even when cardinality estimates are inaccurate. Leis et al.\cite{Leis:QOREALLY:pvldb-2015,Leis:JOB:vldb-2018} conducted experiments on PostgreSQL 9.4, which lacks support for parallel query processing. Given that newer versions of PostgreSQL now include parallel processing features, does Leis's observation remain valid? What are the implications in situations where cardinality estimates are unavailable, i.e., HEAVY.AI? In this paper, using the Join Order Benchmark(JOB), we investigate the gap between the optimal execution plan, determined using actual cardinalities (TrueCard), and the plan generated using cardinality estimates(CE) -- i.e.,  PostgreSQL. As a point of comparison in scenarios lacking cardinality estimates, we introduce a heuristic strategy, denoted as $\text{Simpli}^{2}$ (no-CE), devised to delay foreign key/foreign key joins.

Figure \ref{fig:motivation} depicts the runtime disparities among the TrueCard, CE, and no-CE plans across thread counts ranging from 1 to 5 under non-indexed and indexed configurations. In non-indexed settings, CE and no-CE perform nearly as well as TrueCard. Moreover, we observe that the runtime gap narrows as the thread count increases. This finding suggests that the insights previously offered by Leis et al.\cite{Leis:QOREALLY:pvldb-2015,Leis:JOB:vldb-2018} remain applicable in multi-threaded environments and even when cardinality estimates are absent.

In a single-threaded, indexed environment, the performance disparity between TrueCard and no-CE is notable. This disparity is substantially mitigated when using cardinality estimates, as evidenced in CE's performance. This highlights the rationale behind the prevalent use of single-threaded environments to evaluate the impact of cardinality estimates on query execution time\cite{Han:CEB-github:2021,Han:CEB:2021}. Interestingly, this runtime discrepancy tends to decrease as the number of threads increases.

The results presented in Figure \ref{fig:motivation} challenge our general understanding of the significance of cardinality estimates in query optimization, where performance is expected to be much worse without cardinality estimates. To better understand this result, further analysis is required. Leis et al.\cite{Leis:JOB:vldb-2018,Leis:QOREALLY:pvldb-2015} have previously conducted an exhaustive study on how inaccuracies in cardinality estimates can lead to prolonged query runtimes, emphasizing the need for accurate estimates. Building upon this, Lee et al.\cite{Leis:JOB:vldb-2018,Leis:QOREALLY:pvldb-2015} investigated the phenomenon within an industrial database system, Microsoft SQL Server, demonstrating the effects of progressively introducing larger cardinality errors in varying subsets of query sub-expressions on query optimization.

However,  the question of how a query optimizer functions without any cardinality estimates and what is the performance disparity when optimizing with no cardinality estimates versus true or estimated cardinality remains largely unexplored. In this paper, we aim to shed light on these queries through a comparative analysis. We examine the performance traits of three optimization strategies: no-CE (optimization without Cardinality Estimates), TrueCard (optimization with True Cardinality), and CE (optimization with Cardinality Estimates). Our primary goal is to unravel the following key questions:

\begin{itemize}
    \item  What is the impact of different methods -- no-CE, TrueCard, and CE -- on query cost and execution time? 
	\item How does the performance of the no-CE method compare to that of optimizers used in in-memory databases like DuckDB, MonetDB, and HEAVY.AI?
    \item To what extent are cardinality estimates necessary in environments where indexes are absent?
    \item How do factors such as parallel processing and the selection of physical operators influence query runtime?
\end{itemize}

In an effort to address these questions, we conducted comprehensive experiments utilizing the Internet Movie Database (IMDB) dataset and Join Order Benchmark (JOB) queries. These experiments were carried out across several database systems: PostgreSQL, MonetDB, DuckDB, and HEAVY.AI. For implementing the CE approach, we utilized PostgreSQL's built-in cardinality estimation mechanisms. In contrast, for the no-CE approach, we employed $Simpli^2$\cite{Datta:Simpli2:arXiv-2021}, a heuristic-based query optimizer that operates independently of cardinality estimates. This diverse experimental setup allowed us to thoroughly investigate and compare the performance of different query optimization strategies under varied conditions.

\begin{figure*}[htbp]
	\centering
	\includegraphics[width=0.85\textwidth]{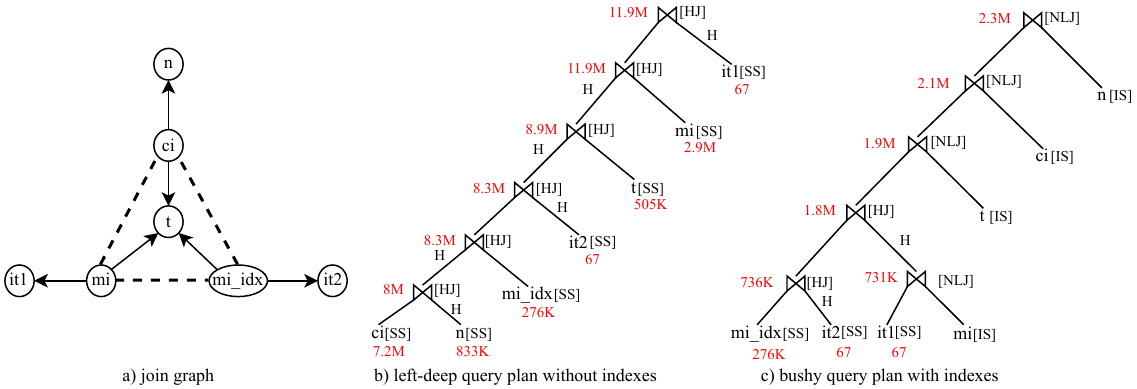}
	\caption{Join graph and two execution plans -- or trees -- for query JOB 18a. Arrows stand for one-to-many joins while dashed lines for many-to-many joins, cost of operations highlighted in red (a). Left-deep query plan with exclusive sequential scan (SS) access (b). Bushy query plan with sequential and index scan (IS) access (c).}
	\label{fig:join-graph+query-plans}
\end{figure*}

The organization of this paper is as follows. In section \ref{sec:query-opt}, we delve into query optimization and its fundamental components by illustrating an example query. Section \ref{sec:postgres} is dedicated to an in-depth discussion of CE query optimization, specifically within the context of non-indexed and indexed settings. Section \ref{sec:noest} delves into the no-CE approach, particularly discussing the $Simpli^2$ algorithm and its implementation details. Section \ref{sec:experiments} outlines our experimental framework and provides an extensive comparative analysis of the no-CE, TrueCard, and CE approach. Sections \ref{sec:related-work} and \ref{sec:conclusions} explore related work and conclude with a summary. The source code and data for our experiments are available on GitHub\cite{Datta:good_plan}.

\section{QUERY OPTIMIZATION}\label{sec:query-opt}

\begin{figure}[htbp]
	\centering
	\includegraphics[width=0.80\textwidth]{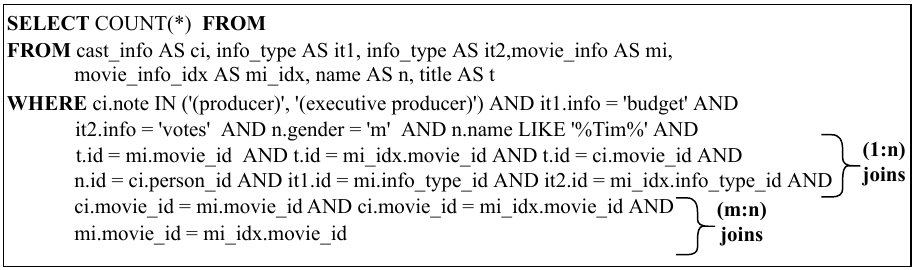}
	\caption{Query 18a from JOB benchmark.}
	\label{fig:sql-query}
\end{figure}

In this Section, we present the query optimization process from PostgreSQL based on the JOB benchmark query 18a, depicted in Figure~\ref{fig:sql-query}. The PostgreSQL optimizer has a standard architecture derived from the traditional cost-based System R query optimizer~\cite{Selinger:APSRDMS:sigmod-1979} and is used as a baseline in this and many other publications~\cite{Leis:QOREALLY:pvldb-2015,Leis:JOB:vldb-2018}.

\noindent
\textbf{Query and join graph.}
Query 18a joins seven tables with nine join predicates -- including a four-table clique among tables \textit{t}, \textit{ci}, \textit{mi}, and \textit{mi\_idx} -- and has four selection predicates on table \textit{ci}, \textit{n}, and the two instances of \textit{it}, respectively. The SQL query can be illustrated as a join graph (Figure~\ref{fig:join-graph+query-plans}a) in which every table is represented as a vertex. Edges represents join predicates connecting the corresponding vertices. For example, the join predicate \textit{ci.movie\_id = t.id} is depicted as the directed arrow between the \textit{ci} and \textit{t} vertices, while the join predicate \textit{ci.movie\_id = mi.movie\_id} as the dashed line between \textit{ci} and \textit{mi}. The first join is a one-to-many ($1:n$) -- or a primary key/foreign key -- join meaning that every tuple from \textit{ci} joins with exactly a single tuple from \textit{t}. The arrow points to the primary key table \textit{t}. The join \textit{ci.movie\_id = mi.movie\_id} is a many-to-many ($m:n$) -- or a foreign key/foreign key -- join. In this case, a tuple from \textit{ci} can join with any number of tuples from \textit{mi} -- including no tuples at all. Likewise, a tuple from \textit{mi} can join with any number of tuples from \textit{ci}. The distinction between these types of joins is important in query optimization because the cardinality of $1:n$ joins is exactly determined by the size of the foreign key table after applying filter predicates, while the cardinality of $m:n$ joins can range from $0$ to the product of the two table sizes $|ci|\cdot|mi|$.

\noindent
\textbf{Query execution plan.}
The query execution plan defines the order in which tables are joined to generate the query result. Execution plans can be represented as a binary tree in which the input tables appear as leaves while the intermediate nodes correspond to joins. At every node, two tables are joined based on their corresponding predicate(s), finishing at the root of the tree. In addition to the join order, the execution plan also determines the access path for every table and the physical join operator. The most common access paths are full table sequential scan (SS) and index scan (IS) with lookup in an unclustered B+ tree\cite{Ullman:db-book}. The join implementations available in PostgreSQL are nested-loop join (NLJ) with or without index lookup, in-memory hash join (HJ), and out-of-memory sort-merge join (SMJ)\cite{postgres}. The query plan is determined during query optimization and is not subsequently modified during execution.

Figure~\ref{fig:join-graph+query-plans} depicts two equivalent query plans for query 18a. Although these plans join the tables in a different order, and contain different table access paths and physical join operators, they produce the same correct query result. The plan in Figure~\ref{fig:join-graph+query-plans}b has a linear structure in which every join has a base table operand. These type of plans are known as left-deep or right-deep trees\cite{Yannis:trees:1991}. All the table access paths are sequential scans (SS) and the only join implementation is the hash join (HJ). The hash table (H) is built either on the base table or on the join result, depending on their estimated sizes -- or cardinality. The plan in Figure~\ref{fig:join-graph+query-plans}c has a bushy structure because it includes a join between two intermediate tables that are the result of previous joins $(mi\_idx \Join it2) \Join (mi \Join it1)$. Since in this setting there are indexes defined on every base table, the access path is an index scan (IS) for each of the tables except \textit{it1}, \textit{it2}, and \textit{mi\_idx}. IS is always coupled with a nested-loop join (NLJ) because this allows for immediate identification of joining tuples.

\subsection{Cost Model}

For query 18a, the goal of the query optimizer is to find the best plan between the plans in Figure~\ref{fig:join-graph+query-plans} and all other equivalent plans. To achieve this, the query optimizer might compare any two query plans without actually executing them. This is realized by assigning a cost to every plan using a cost function -- or cost model -- that can be computed quickly and without accessing the tables. The cost function is defined based on the optimization objective -- which is often the query runtime. Such a cost function has to assign to every plan a cost which is correlated with their execution time -- a plan that has a higher runtime than another should have the larger cost. Since the best indicator for runtime is the total number of processed tuples -- or cardinality -- by the plan operators, the most common cost functions are defined in terms of operator cardinality. The plan cardinality -- the sum of the individual operator cardinalities -- depends both on the join order as well as the physical operators. 

The PostgreSQL cost model distinguishes between accessing a tuple from disk (I/O cost) and processing a tuple in memory (CPU cost). The cost of an operator combines the I/O and CPU costs in a weighted sum, where the weights are constant parameters. While the optimizer defines default values for these parameters, they should be carefully calibrated in order to maximize the correlation between the cost function and runtime -- especially for in-memory databases where page access is much faster.

\begin{figure}[htbp]
	\centering
	\includegraphics[width=0.75\textwidth]{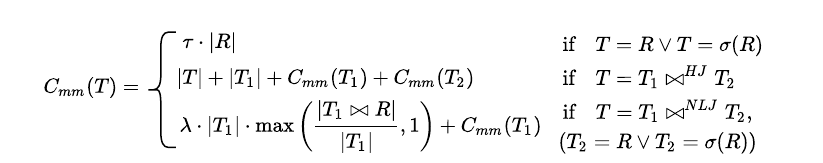}
	\caption{Main-memory cost function defined in~\cite{Leis:JOB:vldb-2018}.}
	\label{fig:in-mem-cost-func}
\end{figure}

Leis et al.~\cite{Leis:JOB:vldb-2018} introduce a simple cost function that is based on operator cardinality for main-memory setups. The cost function $C_{mm}$ is defined recursively for every node in the query tree. The total cost of a node sums the cost of its children with the cost of just itself. Thus, the cost of the complete query plan is obtained at the root of the tree. $C_{mm}$ considers only sequential scans (SS), hash joins (HJ), and index nested-loop joins (NLJ). The cost of SS is the cardinality of the table scaled down by a factor $\tau=0.2$ that quantifies the ratio between scanning a tuple and processing it in a join operator. The cost of HJ sums up the cardinality of the side on which the hash table is built $T_{1}$ with the number of tuples generated by the join operator $T$. The tuples that do not probe successfully are discarded because they incur negligible cost compared to other operations. Finally, the cost of (index) NLJ is dominated by the cardinality of the scanned child $T_{1}$. This value is scaled up by the constant parameter $\lambda=2$ -- assuming that an index lookup is twice as expensive as a hash lookup -- and by the ratio of tuples that have multiple join matches. Notice that parameters $\tau$ and $\lambda$ depend only on the logical algorithms implementing the operator and are independent of physical system characteristics.

Figure~\ref{fig:join-graph+query-plans} labels the cost for two query plans. The cost of the left-deep plan without indexes (b) is 5X larger than the cost of the bushy plan with indexes (c). The main reason for this is because index scans (IS) incur negligible costs while sequential scans (SS) have a cost proportional to the table cardinality. For example, SS($ci$) contributes a cost of 7.2M to (b) while IS($ci$) incurs no cost in (c). The ratio between the cost of NLJ and HJ is at most a factor of $\lambda$. Since join sizes are much smaller than table sizes, the additional cost incurred by NLJ over HJ is much smaller compared to the SS cost. The difference in the cost between (b) and (c) is confirmed by their runtime in PostgreSQL where (c) runs 3X faster than (b).

\subsection{Cardinality Estimation}
Join cardinality serve as the primary input to a cost function. However, obtaining exact join cardinality requires performing the join itself. Query optimization aims to determine the optimal join order without actually executing the joins. Cardinality estimation -- predicting the number of rows returned by an operation -- addresses this challenge by approximating join cardinalities, providing the cost function with estimates to compute the plan cost.

Query optimizers utilize various techniques to estimate the cardinality of join operators, which are vital for evaluating the plan's cost and choosing the most efficient one. Common methods include histograms, sampling, sketches, and, recently, machine learning models\cite{Han:CEB:2021}. Histograms are a widely used technique for estimating value distributions within attributes, but struggle with capturing correlations across join attributes. Sampling, which involves estimating cardinality based on a random data sample, is advantageous when exact statistics computation is infeasible due to dataset size or complexity. However, this method may produce less accurate estimates, especially for skewed data distributions. Sketches offer a space-efficient, scalable solution for approximating large dataset cardinalities or complex operations but require domain expertise to balance accuracy and performance. Machine learning models have been used recently to predict cardinalities based on observed data and query patterns. While effective for single-table cardinality estimation, these models may struggle with complex join conditions and may require significant training time \cite{han2021cardinality}.

Cardinality estimates often exhibit inaccuracies due to discrepancies between the assumptions made by the cardinality estimator and the actual data distribution\cite{Leis:QOREALLY:pvldb-2015, Leis:JOB:vldb-2018}. For instance, the PostgreSQL cardinality estimator presumes uniform value distribution, independence among attributes within a table, and overlapping join keys. However, real-world datasets frequently exhibit non-uniform distributions, inter-attribute correlations within the same table, and potentially non-overlapping join keys. These discrepancies between the true dataset distribution and the cardinality estimator's assumptions results in errors. As the complexity and volume of the data increases, these errors become more unpredictable.

\subsection{Plan Enumeration}

A query plan consists of an ordered sequence of tables -- a join order -- and a selection of physical operators. Plan enumeration iterates through semantically equivalent plans in two search spaces: the \textit{algebraic search space} and \textit{method-structure search space} \cite{Yannis:ACM:journal-1996}. The algebraic search space encompasses all valid plans, or combinations of relational algebra operations, capable of correctly answering the query. For instance, Figure \ref{fig:join-graph+query-plans} depicts two plans from (b) the left-deep search space and (c) bushy search space. Finding an optimal plan is NP-complete \cite{Ibaraki:NP:tds-1984}, and the number of potential plans is factorial with respect to the number of tables. In the left-deep search space alone there are $n!$ different join orders to choose from, including cross joins, where $n$ is the number of tables. The total algebraic search space is even larger, accounting for right-deep, zig-zag, and bushy search spaces.

Upon determining the join order in the algebraic search space, the enumeration process selects suitable physical operators within the method-structure search space. This step involves identifying optimal algorithms for various operations, such as joins (e.g., hash join, merge join, nested loop join), scan and sort operators, and considering the availability of indexes. Here, appropriate physical operators are vital to optimizing runtime.

The plan search space can be explored exhaustively~\cite{Moerkotte:DP:sigmod-2008,Moerkotte:TDDP:pvldb-2013} or heuristically ~\cite{Steinbrunn:HRO:jvldb-1997}. To exhaustively explore the search space, the query optimizer need time and access to join cardinality estimates for all join subsets, while heuristic methods present a trade-off between enumeration time and the number of required cardinality estimates. By constraining the search space, query optimizers examine a smaller subspace. Hybrid approaches, such as those presented by Cai et al. \cite{Cai:PCETUB:sigmod-2019} and Hertzschuch et al. \cite{Hertzschuch:simplicity:cidr-2021}, combine heuristic and exhaustive search while utilizing functional dependency information or the join graph structure derived from SQL queries, respectively. However, limiting the search space raises the risk of selecting a suboptimal plan.

Furthermore, any enumeration algorithm may be misled by inaccurate cardinality estimations, resulting in suboptimal plans. Even minor underestimations and overestimations can mislead the plan enumeration algorithm.

\section{CE Query Optimization}\label{sec:postgres}

CE query optimization utilizes cardinality estimates to guide optimization decisions. This method employs various data synopses and statistics to identify the most efficient plan for executing a query. The optimizer leverages these cardinality estimates to make informed decisions regarding join order, physical operators, and indexes. For instance,  if the optimizer estimates that a specific join would generate a large number of rows, it may opt for a different join order to minimize the plan cardinality and improve performance. Likewise, during index selection, if the optimizer anticipates that a particular index will be highly selective and yield fewer rows, it may favor that index to enhance performance. Accurate cardinality estimation has the potential to boost query performance by allowing the optimizer to make better-informed decisions when executing a query. However, success depends on the precision of the statistics and data synopses.

\begin{figure}[htbp]
	\centering
	\includegraphics[width=0.50\textwidth]{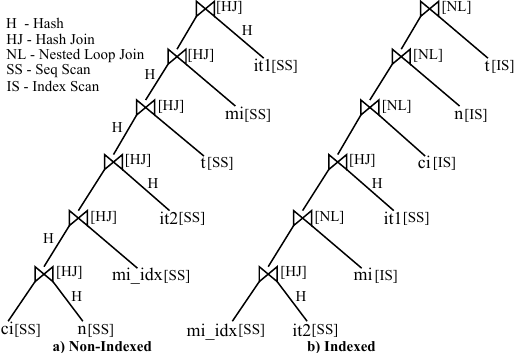}
	\caption{PostgreSQL execution plans for query 18a.}
	\label{fig:pg_noidxvsidx}
\end{figure}

\subsection{Non-Indexed}

Many modern relational databases, such as \textbf{PostgreSQL}, employ a CE query optimizer\cite{postgres}. The PostgreSQL optimizer leverages data and system statistics to estimate the cost of all potential join orders or plans, ultimately selecting the one with the lowest cost. Although the enumeration can benefit from join graphs by traversing only existing edges, the edge type — either primary/foreign key or foreign/foreign key — is not a primary consideration. Redundant and suboptimal orders are efficiently managed through dynamic programming and early pruning techniques\cite{postgres}.

A plan's cost is calculated by aggregating the cost of individual plan operators. The cost of an operator is contingent upon the number of accessed pages and processed tuples, with these quantities weighted by the configurable system parameters, \texttt{seq\_page\_cost} and \texttt{cpu\_tuple\_cost}, respectively\cite{postgres}. Exact cardinalities are only known for base tables, necessitating estimations for all other operators without executing them. This constitutes the cardinality estimation problem in PostgreSQL\cite{postgres}.

Pre-computed data synopses or statistics are utilized for this purpose. PostgreSQL statistics are per-attribute and includes ranges, heavy hitters, the number of distinct values, and equi-depth histograms. Operator costs are estimated by integrating these statistics into formulas that assume uniformity, independence, and inclusion regarding the data\cite{Leis:QOREALLY:pvldb-2015,Leis:JOB:vldb-2018}. Inaccurate estimates result whenever these assumptions are violated, often in queries involving many-to-many joins or multiple predicates.

We execute JOB query 18a in PostgreSQL. To determine the optimal plan, PostgreSQL's enumeration algorithm exhaustively examines all potential join orders of the $ci, n, mi\_idx, it2, t, mi, it1$ tables and the available physical operators. For example, it evaluates permutations such as ($ci \Join n \Join mi\_idx \Join it2 \Join t \Join mi \Join it1$), ($(ci \Join n) \Join (mi\_idx \Join it2) \Join t \Join mi \Join it1$), and others. In the non-indexed configuration, due to our main-memory setup, we constrain the join operators to hash join.

The enumeration algorithm also investigates various options for building and probing hash tables. The optimal plan for query 18a is illustrated in Figure \ref{fig:pg_noidxvsidx}a, where the optimizer chooses to construct a hash table on the join result of ($ci \Join n$) rather than on $mi\_idx$. Subsequently, however, it builds the hash table on the base table $it2$ instead of the join result of ($ci \Join n \Join mi\_idx$). This choice is made because, in certain cases, constructing a hash table on an intermediate join result can be more cost-effective than doing so on a base table.

\subsection{Indexed}

In the presence of indexes, the PostgreSQL optimizer follows a process akin to the one employed in non-indexed settings. The integration of indexes within a database system broadens the choices of available physical operators, as it offers an alternative method for data access beyond a full table scan. Instead of examining all records in a table to locate the desired information, the index facilitates direct access to pertinent data, thereby improving efficiency and overall performance.

Consider JOB query 18a with the optimal indexed join order: $mi\_idx\Join it2\Join mi\Join it1\Join ci \Join n\Join t$ which differs from the non-indexed setting. In the presence of indexes, the optimizer determines if using an index will be beneficial, given that the cost of an index scan differs from that of a sequential scan. For example, if an index is available on the $mi\_idx$ table's join column ($info\_type\_id$) with $it2$ tables join column ($id$), PostgreSQL's cost function evaluates the cost of the join with the index (20K) and without the index (16K) and chooses the more cost-effective option. However, the cost of join operations between $mi$ and $mi\_idx$ tables is different. The cost of the join using an index is 8K, whereas forgoing the index incurs a cost of 234K. Thus, the optimizer chooses the indexed operation.

\section{No-CE Query Optimization}\label{sec:noest}
Inaccurate join cardinality estimation can considerably influence the execution time of a query. If the estimate is underestimated, the optimizer might select a suboptimal join algorithm, such as a nested loop join, which exhibits slow performance for large datasets. Conversely, overestimating could lead the optimizer to opt for a more time-efficient join algorithm, like a hash join, while unnecessarily allocating excessive memory and other resources. This overallocation results in squandered resources and potentially longer execution times due to excessive sorting or hashing. Such issues are predominantly prevalent in queries involving sizeable joins, where the estimation error can be substantial and prompt the query optimizer to adopt a suboptimal plan. Consequently, the query may run for an extended execution time and even timeout.

In contemporary main-memory databases, advanced mechanisms like indexing, caching, and parallel processing enable rapid and efficient execution of join operations. Interestingly, many of these systems achieve impressive performance even in the absence of sophisticated cardinality estimator\cite{heavy-ai, duckdb}. This could be attributed to advancements in hardware and storage technologies, which may have reduced the need for precise cardinality estimates in optimizing join operations in main-memory systems.

\subsection{Simpli-Squared}

\textbf{$\text{Simpli}^{2}$}, is designed to completely eschew the use of synopses, statistics, or cardinality estimations. Instead, it employs the query's join graph and key/foreign key constraints to guide its decision-making. While the incorporation of key/foreign key constraints in join optimization has precedents\cite{Hertzschuch:simplicity:cidr-2021,Cai:PCETUB:sigmod-2019}, those prior approaches have typically combined such constraints with cardinality estimates,  which is not the case here. Additionally, $\text{Simpli}^{2}$ takes table sizes as auxiliary input parameters. These inputs are used to annotate the vertices (tables) and edges (joins) of the join graph. Table vertices are annotated with their respective sizes, while join edges are classified as either one-to-many (1:n) or many-to-many (n:m) based on their key/foreign key constraints.

\DontPrintSemicolon 
\begin{algorithm}
	\caption{Simpli-Squared ($\text{Simpli}^{2}$)}\label{alg:the_alg}
	\KwData{Query join-graph $G(V,E)$ where edges are key/foreign key constraints}
	\KwResult{Join-order $JO$, initialized to an empty list}

	$FK$ $\leftarrow$ list of all foreign key tables from $V$ \\

	Sort $FK$ in ascending order by $C_{S2}$ \\
	\For{$f \in FK$}
	{
		
		$FK\_C$ $\leftarrow$ list of primary key tables adjacent to $f$ \\\
		Sort $FK_c$ in ascending order by cardinality \\
		Append $f$ to $JO$ \\
		\For{$c \in FK_c$ }
		{
			\If{$c \notin JO$} 
			{
				Append $c$ to $JO$
			}
		}
		note : each $f \in FK$ produces separate sub-query after it joins one or more primary key table  \\ 
	}

	\While{$\exists$ table c $\in V \ni c \notin$ $JO$}
	{
		\If {$\exists  c' \in JO \ni (c,c') \in E $}
		{
			Append $c$ to $JO$
		}
	}
\end{algorithm}

\begin{figure*}[htbp]
	\centering
	\includegraphics[width=.99\textwidth]{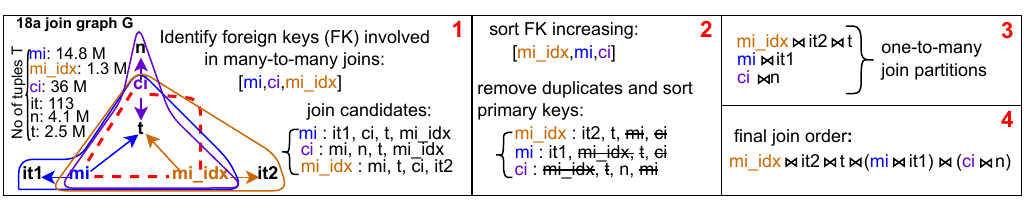}
	\caption{$\text{Simpli}^{2}$ example on JOB query 18a. Key/foreign key constraints are represented by arrows pointing to the key attribute. Many-to-many joins are depicted with dashed lines. The join candidates of a table are grouped into polygons.}
	\label{fig:alg-18a}
\end{figure*}

$\text{Simpli}^{2}$ addresses join order as a graph problem, with the goal of identifying the optimal traversal path, which is determined incrementally starting from a source vertex. The traversal of subsequent vertices is guided by a set of heuristics aimed at minimizing join size. Firstly, only edges from included vertices are considered, thereby eliminating cartesian products. Secondly, One-to-many joins are prioritized over many-to-many joins, as their sizes are more predictable -- only as large as the table containing the foreign key. Many-to-many joins are incorporated only when necessary to avoid cartesian products. Lastly, the number of tuples serves as the sole criterion for choosing between tables, with smaller tables taking precedence over larger ones.

These heuristics can be intuitively understood in the context of the JOB schema. Many-to-many joins typically involve two large fact tables, while one-to-many joins occur between a fact table and a significantly smaller table. By prioritizing one-to-many joins, the number of rows participating in the join from the fact tables is reduced. Furthermore, prioritizing smaller tables decreases the likelihood of generating large join results early in the join order.

$\text{Simpli}^{2}$ has two optimization objectives: minimizing the number of accessed tuples, and maximizing the utilization of available indexes. The cost of joining a foreign key table is determined by its cardinality when indexes are unavailable. When there are indexes, its cost is inversely proportional to the number of adjacent primary key tables. We express these objectives in the following formula
\[
C_{S2}(FK_m)  = \left.
  \begin{cases}
    |FK_m|, & \text {non-indexed} \\
    \frac{|FK_m|} { 2^{T_{FK_m}}}, & \text {indexed} \\
  \end{cases}
  \right.
\]
Where $FK_m$ is a foreign key table and $T_{FK_m}$ is the number of primary key tables adjacent to $FK_m$. The cost of $FK_m$ is calculated differently depending on the presence of indexes. The join order of a query is determined by sorting it's foreign key tables in ascending order of their cost. 

\begin{figure}[htbp]
	\centering
	\includegraphics[width=0.50\textwidth]{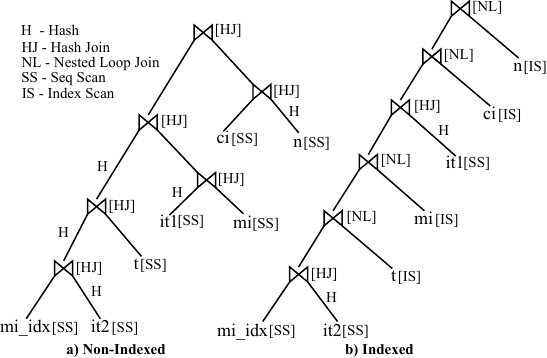}
	\caption{$Simpli^2$ execution plans for query 18a.}
	\label{fig:s2_noidxvsidx}
\end{figure}

\subsection{Non-Indexed} 
The $\text{Simpli}^{2}$ algorithm is outlined in Algorithm~\ref{alg:the_alg}. It starts by identifying the foreign key tables and their primary key join candidates, splitting the join graph into multiple -- possibly overlapping -- components. The join order is determined by sorting the foreign key tables using $C_{S2}$, their cardinalities, in ascending order. Each foreign key table is appended to the join order, followed by its component. The primary key tables in each component is appended to the join order in ascending order of their cardinality. If the same primary key table is present in multiple components, it will only be appended to the join order with the component where it initially appears. Any remaining tables not yet part of the join order are subsequently appended. Each component in the join graph is treated as a subquery, with the exception of the first one, resulting in a bushy join order.

We execute JOB query 18a with the $\text{Simpli}^{2}$ algorithm. We compute the components for each foreign key table: $mi\_idx$, $mi$, and $ci$. The components are then combined, resulting in the join order depicted in Figure \ref{fig:s2_noidxvsidx} as a bushy tree.

\subsection{Indexed} 
In the presence of indexes, assuming each table has at least one index, the $\text{Simpli}^{2}$ algorithm has the additional objective of maximizing the utilization of those indexes. While the algorithm itself remains the same, the cost determined by $C_{S2}$ differs from the non-indexed setting. To calculate the cost of a foreign key table with indexes, $C_{S2}$ divides the table's cardinality $|FK_m|$ by $2^{T_{FK_m}}$ where $T_{FK_m}$ represents the number of adjacent primary key tables. Thus, foreign key tables with more adjacent primary key tables are prioritized in the join order. Bushy trees, which minimize access to base tables later in the join order and subsequently reduce index usage, are counterproductive to our objective. To address this, we default to left-deep trees when there are indexes. The impact of these modifications on query 18a's join order is depicted in Figure \ref{fig:s2_noidxvsidx}. Additionally, the choice of physical operators is adapted to benefit from indexes. For instance, index scans are preferred over sequential scans, and indexed nested loop joins are preferred over hash joins.

Following Hertzschuch et. al ~\cite{Hertzschuch:simplicity:cidr-2021}, we rewrite queries using the explicit join order determined by our algorithm. In a non-indexed setting, we construct a subquery for each join component -- except the first. When executing rewritten query with a database system, e.g., PostgreSQL, the optimizer must be configured to follow the given join order and disable subquery unnesting ~\cite{Hertzschuch:simplicity:cidr-2021}. We leave the selection of physical operators up to the optimizer's default, which may use statistics as is the case when indexes are available.

\section{EXPERIMENTAL EVALUATION}\label{sec:experiments}

For our experiments, we follow Cai et al.\cite{Cai:PCETUB:sigmod-2019,Cai:pessimistic-github:2021} and inject user-provided cardinalities of subqueries into PostgreSQL database system version 14.2. PostgreSQL is open-source, which allows researchers to access optimization, and execution parameters for their experiments. Additionally, PostgreSQL has a wide range of features, such as support for various data types and query optimization techniques, which make it well-suited for our experiment. 

We configure PostgreSQL for a main-memory setup, in line with prior work \cite{Leis:JOB:vldb-2018,Leis:QOREALLY:pvldb-2015}. Specifically, by increasing \texttt{work\_mem} from 4 megabyte to 128 gigabyte, \texttt{shared\_buffers} from 32 megabyte to 128 gigabyte, and \texttt{effective\_cache\_size} from 4 gigabyte to 128 gigabyte. We also increase the \texttt{geqo}\texttt{\_threshold} parameter to 18 joins, which determines the threshold for switching from dynamic programming to heuristic search. Lastly, we set \texttt{from\_collapse\_limit} and \texttt{join\_collapse\_limit} to 1 for $\text{Simpli}^2$ queries, so that the query optimizer does not reorder joins or unnest subqueries.

\begin{figure*}[htbp]
	\centering
	\includegraphics[width=.75\textwidth]{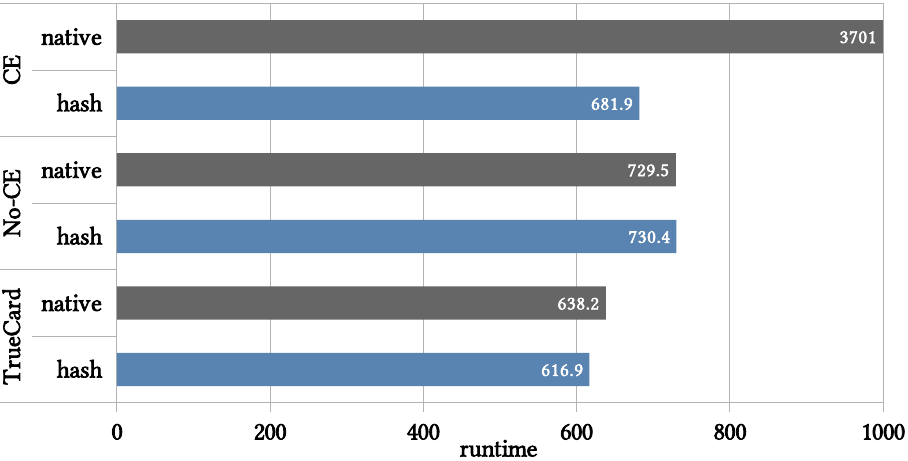}
	\caption{Cumulative Runtimes(seconds) for Join Order Benchmark Queries: Comparison of Native and Hash-Only Join Operators in a Non-Indexed setting}
	\label{fig:hash_join_motivation}
\end{figure*}

We conduct experiments on both indexed and non-indexed configurations. Indexes are built on all primary key and foreign key columns following JOB\cite{JOB-github}. In the indexed setting, we disable sort-merge join, while in the non-indexed setting, sort-merge and nested-loop join are both disabled. Hash joins are preferred in main-memory and non-indexed setups to ensure optimal performance\cite{Leis:QOREALLY:pvldb-2015,Leis:JOB:vldb-2018}.

In our experiment, we also include tests on DuckDB\cite{duckdb} and HEAVY.AI\cite{heavy-ai}, both of which operate as main-memory database systems. Furthermore, we extend our experiments to MonetDB\cite{monetdb-web}, a database system that employs a columnar storage architecture.

\noindent
\textbf{PosrgreSQL non-indexed}

In the absence of indexes, this study evaluates the performance of CE, No-CE, and TrueCard under two distinct configurations of PostgreSQL: (a) the native setting, where PostgreSQL's query planner is free to select from Hash Join, Merge Join, and Nested Loop Join operators, and (b) the hash-enforced setting, where the query planner's choices are restricted to always use Hash Join by disabling Merge Join and Nested Loop Join options. The outcomes are illustrated in Figure \ref{fig:hash_join_motivation}, where the X-axis quantifies the cumulative runtime for the Join Order Benchmark workload. On the Y-axis, the comparison focuses on the CE, No-CE, and TrueCard, with each method's performance outlined by a pair of bars representing the runtimes for the 'native' and 'hash' configurations, respectively. The runtimes are recorded in seconds, and a query timeout threshold is established at 300 seconds.

\begin{table}[htbp]
	\centering
	\begin{tabular}{@{}lrrr@{}}
	\toprule
	& \textbf{Hash Join (HJ)} & \textbf{Merge Join (MJ)} & \textbf{Nested Loop (NL)} \\ \midrule
	CE       & 756                     & 4                        & 104                       \\
	No-CE    & 847                     & 13                       & 4                         \\
	TrueCard & 842                     & 5                        & 17                        \\ \bottomrule
	\end{tabular}
	\caption{Number of different join operators in native Non-Indexed setting}
	\label{tab:joinmethods}
\end{table}

In the native configuration, the CE approach demonstrated the longest runtimes among the methods evaluated, being approximately fivefold and 5.8-fold slower than the No-CE and TrueCard approaches, respectively, thus marking it as the least efficient method. This inefficiency is further underscored by six queries that reached the timeout limit of 300 seconds. For cumulative runtime calculation, these timeouts contribute 300 seconds each to the total runtime of 3701 seconds. The reasons for this extended duration are detailed in Table \ref{tab:joinmethods}. Within the scope of the Join Order Benchmark (JOB) workload, which encompasses a total of 864 join operations, the CE method opted for Nested Loop Joins in 104 instances. This frequency is substantially higher compared to those chosen by the No-CE and TrueCard methods. The preference for Nested Loop Joins is attributed to the underestimation of join cardinalities.

When the PostgreSQL settings are adjusted to utilize Hash Joins exclusively, we observe a substantial improvement in runtime, with an 81\% reduction to 681.9 seconds for the entire workload. Furthermore, this Hash Join-exclusive arrangement yields a more consistent performance and mitigates the potential for selecting suboptimal join operators as a consequence of cardinality estimation errors. Owing to these benefits, this setting is adopted as the standard for non-indexed settings in the remainder of the paper.

\noindent
\textbf{Hardware} Each database is configured within its own Ubuntu 20.04 LTS Docker container. These containers are deployed on a machine featuring an Intel Xeon E5-2660 v4 (2.00GHz) processor with 28 CPU cores, 256 GB of RAM, and HDD storage.  

\noindent
\textbf{Dataset} Many prior works on query processing and optimization use standard benchmarks like TPC-H, TPC-DS, or the Star Schema Benchmark (SSB)\cite{Boncz:TPC-H,ONeil:star-schema,Poess:TPC-DS}. However, Leis et al. \cite{Leis:JOB:vldb-2018} argue that, while these benchmarks serve well in assessing the performance of query engines, they are not good for evaluating query optimizers. We use JOB (Join Order Benchmark)\cite{JOB-github} derived from the IMDB (Internet Movie Database) dataset\cite{Boncz:imdb-data}, a widely recognized benchmark for assessing query optimizer performance in real-world scenarios. JOB consists of 113 queries of varying complexity with up to 28 join predicates. The IMDB dataset features skewed attributes and cross-table correlations\cite{Leis:JOB:vldb-2018}.

\subsection{Analysis between CE and no-CE}\label{ssec:analysis-ceVSnoce}
\begin{figure*}[htbp]
	\centering
	\includegraphics[width=\textwidth]{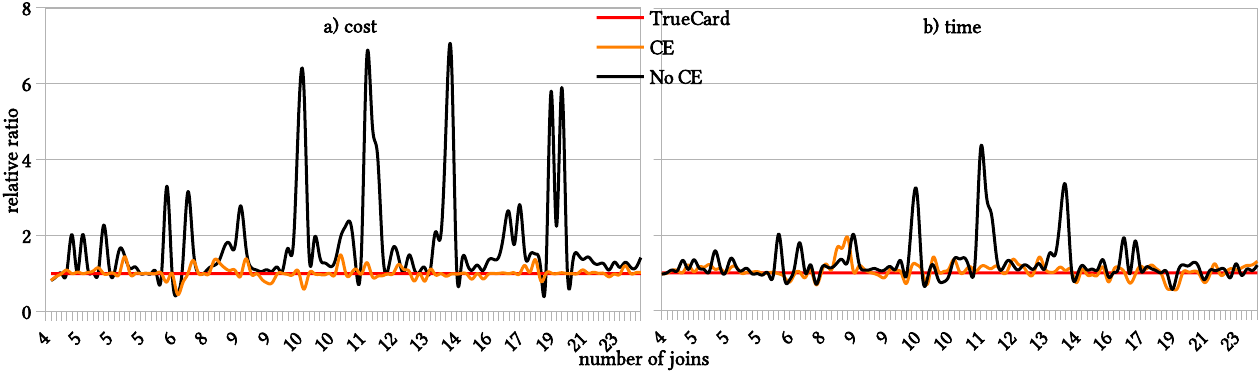}
	\caption{Cost and time [non-indexed setting] --- as a normalized ratio to TrueCard.}
	\label{fig:noidx_analysis}
\end{figure*}

\begin{figure*}[htbp]
	\centering
	\includegraphics[width=\textwidth]{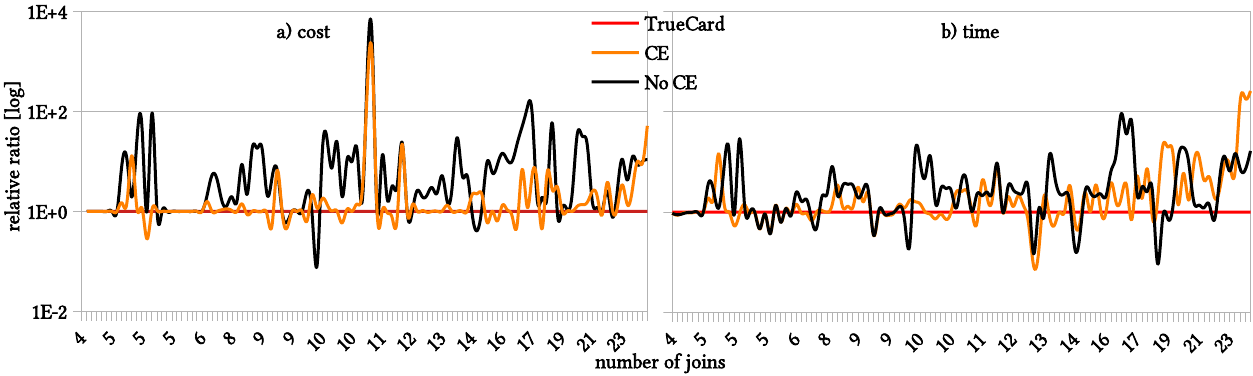}
	\caption{Cost and time [indexed setting] --- as a normalized ratio to TrueCard.}
	\label{fig:fkidx_analysis}
\end{figure*}

In this section, we conduct a comparative analysis of TrueCard, CE, and no-CE within the PostgreSQL framework, focusing on both cost and runtime. Join orders from no-CE are integrated into the PostgreSQL system via query rewriting. Exact cardinalities for TrueCard are collected by executing all subqueries for each query, which are then injected during query runtime. The cost of each query is calculated using the cost function defined in \cite{Leis:JOB:vldb-2018}, which is discussed in Section \ref{sec:query-opt}. Although PostgreSQL's default cost function was an option, we opted for Leis's \cite{Leis:JOB:vldb-2018} cost function due to its suitability for main-memory setups focused on operator cardinality. Query runtime was collected by running each query 11 times and taking its median. These results are illustrated in Figures \ref{fig:noidx_analysis} and \ref{fig:fkidx_analysis}.

\noindent
\textbf{Non-Indexed - } We evaluate the costs of JOB plans without indexes for CE, no-CE, and TrueCard, with the cost ratio relative to TrueCard depicted in Figure \ref{fig:noidx_analysis}a. The cost of CE plans are usually similar to TrueCard's. Even in the worst case, they are negligibly up to 1.5 times higher. The cost of no-CE plans is also similar to TrueCard's, except for a few instances where they can be as much as 6.5 times higher. This is due to the bushy join structure of no-CE method, which generates more intermediate join results compared with a linear structure, i.e., left-deep. In contrast, the join orders utilized by TrueCard and PostgreSQL are primarily left-deep, minimizing the production of join results.

While the cost of a query plan can indicate its execution time, it is not always an accurate predictor. Factors like database configuration and available resources also influence runtime but may not be accurately represented in the cost model. The cumulative cost ratio of JOB plans between CE and TrueCard is 1.01, while the runtime ratio is 1.08 -- which is 8\% slower than TrueCard. Conversely, the cost ratio between no-CE and TrueCard is 1.69, yet the runtime ratio is 1.28 -- 28\% slower than TrueCard. Overall, the cost function tends to overestimate.

\noindent
\textbf{Indexed - } Figure \ref{fig:fkidx_analysis}a depicts the impact of indexing on query plan costs. In the most unfavorable scenarios, the cost of plans using cardinality estimates (CE) can escalate to as much as 1370 times greater than those using TrueCard's exact cardinalities. However, the ratio of these costs exhibits considerable variability across different plans, more so than in non-indexed settings. While some plans do incur elevated costs, the aggregate workload cost for plans using CE is merely 1.13 times that of TrueCard. This is primarily due to the majority of plans having costs similar to, or occasionally even lower than, TrueCard's, thus counterbalancing the effects of costlier outliers. No outliers were observed regarding query runtime, and the cumulative workload runtime for CE-based plans is 1.88 times that of TrueCard's.

In extreme scenarios, no-CE plans can incur costs that are up to 3,000 times greater than TrueCard. Despite being optimized for maximum index utilization, the absence of cardinality estimates often leads to suboptimal query plans. This manifests as a cumulative workload cost that is 4.05 times higher when using no-CE, compared to TrueCard. Interestingly, the no-CE approach exhibits runtimes that are 3.02 times longer than those of TrueCard. However, for certain complex queries involving multiple joins, no-CE surprisingly outperforms TrueCard in terms of runtime. This advantage is likely due to the relatively low optimization time required by no-CE, even for complex, join-heavy queries. In contrast, plans utilizing cardinality estimates (CE) and TrueCard necessitate additional time for optimization, consequently affecting query runtime.  

\begin{table}[htbp]
	\centering
	\begin{tabular}{@{}lllrr@{}}
		\toprule
			& & \textbf{TrueCard (ratio)} & \textbf{CE (ratio)} & \textbf{no-CE (ratio)}\\ \midrule
			\multirow{2}{*}{non-indexed}  & cost(million) & 994.8 (1)  & 1006.6 (1.01) & 1680.4 (1.68)  \\
			 & runtime(seconds) & 324.6 (1)  & 349.4 (1.07) & 416.3 (1.28)  \\ \midrule
			 \multirow{2}{*}{indexed}  & cost(million) & 257.03 (1)  & 290.02(1.13) & 1041.6(4.05)  \\
			 & runtime(seconds) & 102.3 (1)  & 192.5 (1.88) & 309.2(3.02)  \\ \bottomrule
	\end{tabular}
	\caption{Summary of Aggregated Costs (in Millions) and Runtime (in Seconds) (Join Order benchmark). Ratio relative to TrueCard values}
	\label{tab:agg_cost_runtime}
\end{table}

\begin{table}[htbp]
	\centering
	\begin{tabular}{@{}lrr@{}}
		\toprule
			& \textbf{simplicity} & \textbf{no-CE}\\ \midrule
			non-indexed & 271     & 295           \\ 
			indexed     & 2163    & 308           \\ \bottomrule
	\end{tabular}
	\caption{Summary of Simplicity and no-CE method runtime in both non-indexed and index setting}
	\label{tab:simplicity_res}
\end{table}

\noindent
\textbf{Simplicity - } We compare no-CE method with Hertzschuch et al.'s Simplicity algorithm \cite{Hertzschuch:simplicity:cidr-2021}, from which the no-CE approach is derived. The key distinction between Simplicity and no-CE lies in their handling of estimates: while Simplicity utilizes PostgreSQL estimates to compute an upper bound, no-CE abstains from using any such estimates. In a non-indexed environment, the no-CE method exhibits an 8\% performance degradation relative to Simplicity. Conversely, in an indexed setting -- Simplicity is not optimized for taking benefit from indexes -- no-CE outperforms Simplicity by a factor of 7.

\subsection{Optimization Time}\label{ssec:opt-time}

Query runtime consists both optimization and execution time. Optimization time involves the process of determining the most efficient execution plan for a query. An extended optimization time can result in an overall longer query runtime, even if execution time is brief. In contrast, reduced optimization time may decrease overall query runtime by enabling earlier query execution. It is important to note that optimization time does not directly correlate with query runtime. Following optimization, the query is executed.

\begin{figure*}[htbp]
	\centering
	\includegraphics[width = \textwidth]{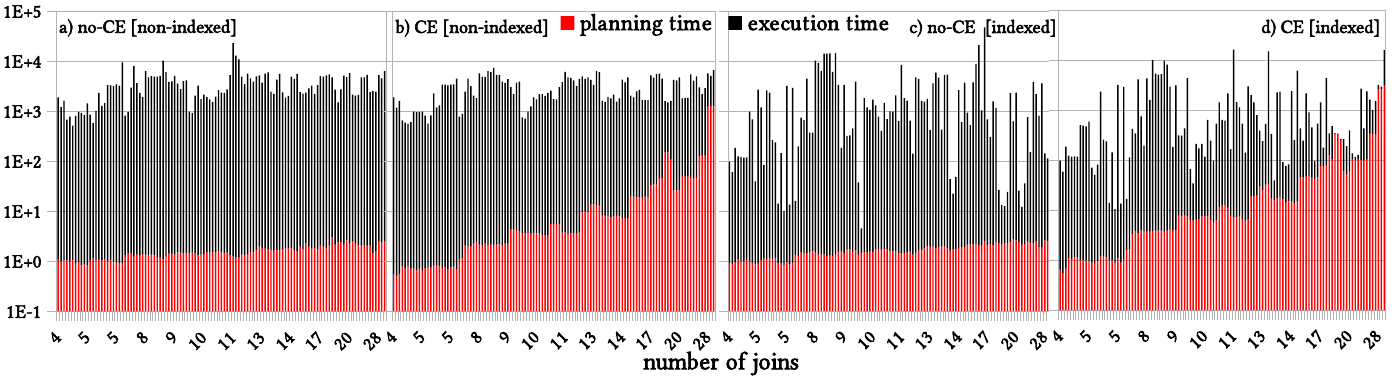}
	\caption{Optimization and execution time in non-indexed and indexed setting.}
	\label{fig:OptVsExecTime}
\end{figure*}


To investigate the impact of optimization time in CE and no-CE, we experiment on both non-indexed and indexed settings. Figure \ref{fig:OptVsExecTime} displays the results, with queries sorted by the number of joins on the X-axis and runtime on the Y-axis.

\noindent
\textbf{Non-Indexed - }Figure \ref{fig:OptVsExecTime}a,b illustrates optimization and execution time for JOB in non-indexed settings where queries are sorted based on the number of joins -- there might be multiple queries that share the same number of joins. While there is no substantial difference in runtime between CE and no-CE for all JOB queries, their optimization time varies. In the case of no-CE, optimization time remains relatively consistent across JOB queries, regardless of their complexity. In contrast, CE optimization time increases with query complexity. 

Since no-CE uses a deterministic greedy algorithm to determine join order, exploration is minimal, resulting in low optimization time. CE, on the other hand, utilizes an exhaustive search algorithm. As the number of joins grows, the search space expands, leading to increased optimization time. As demonstrated in Figure \ref{fig:OptVsExecTime}b, CE's optimization time significantly grows with the number of joins. Despite the substantial difference in optimization time between CE and no-CE, the overall query runtime are similar in the non-indexed setting.

\noindent
\textbf{Indexed - } With indexes, the optimization time for both CE and no-CE methods remains largely similar to the non-indexed setting. For CE, as the number of joins increases, the search space grows, and this expansion becomes more pronounced due to the availability of indexes. In certain cases, we observe that the optimization time for complex queries in CE surpasses no-CE's query runtime. This suggests that the benefits of optimization in CE begins to diminish as the number of joins increases.

To address this issue, traditional databases, i.e., PostgreSQL, limit the exhaustive search algorithm for queries involving more than 12 tables by default and employ a heuristical genetic algorithm for optimization past this threshold. Following Leis et al. \cite{Leis:JOB:vldb-2018,Leis:QOREALLY:pvldb-2015}, we increase this threshold to 18 tables in our experiments, meaning it will never be exceeded.

\subsection{Good Plans Despite Bad/No Cardinalities}\label{ssec:no-ce-no-idx}

Table \ref{tab:agg_cost_runtime} presents an aggregated analysis of costs and runtimes for the Join Order Benchmark, comparing the TrueCard method with CE and no-CE in indexed and non-indexed settings. In non-indexed settings, the CE and no-CE costs are 1.01 and 1.68 times that of TrueCard, respectively, while runtimes are 1.07 and 1.28 times greater. Despite the potential inaccuracies in cardinality estimation and the absence of any estimates in the no-CE approach, the performance of CE and no-CE closely aligns with the TrueCard baseline. Moreover, the performance disparity among the methods is less pronounced in non-indexed settings compared to indexed ones. A detailed examination of Figure \ref{fig:noidx_analysis} reveals that, for the majority of queries, all methods under comparison perform similarly, barring a few outliers. Contrary to expectations that inaccuracies in CE and the absence of cardinality estimates in no-CE would lead to sub-optimal execution plans and extended runtimes, our observations do not substantiate this in most cases. The subsequent sections delve into the underlying reasons for this observation.

\begin{figure}[htbp]
	\centering
	\includegraphics[width = \textwidth]{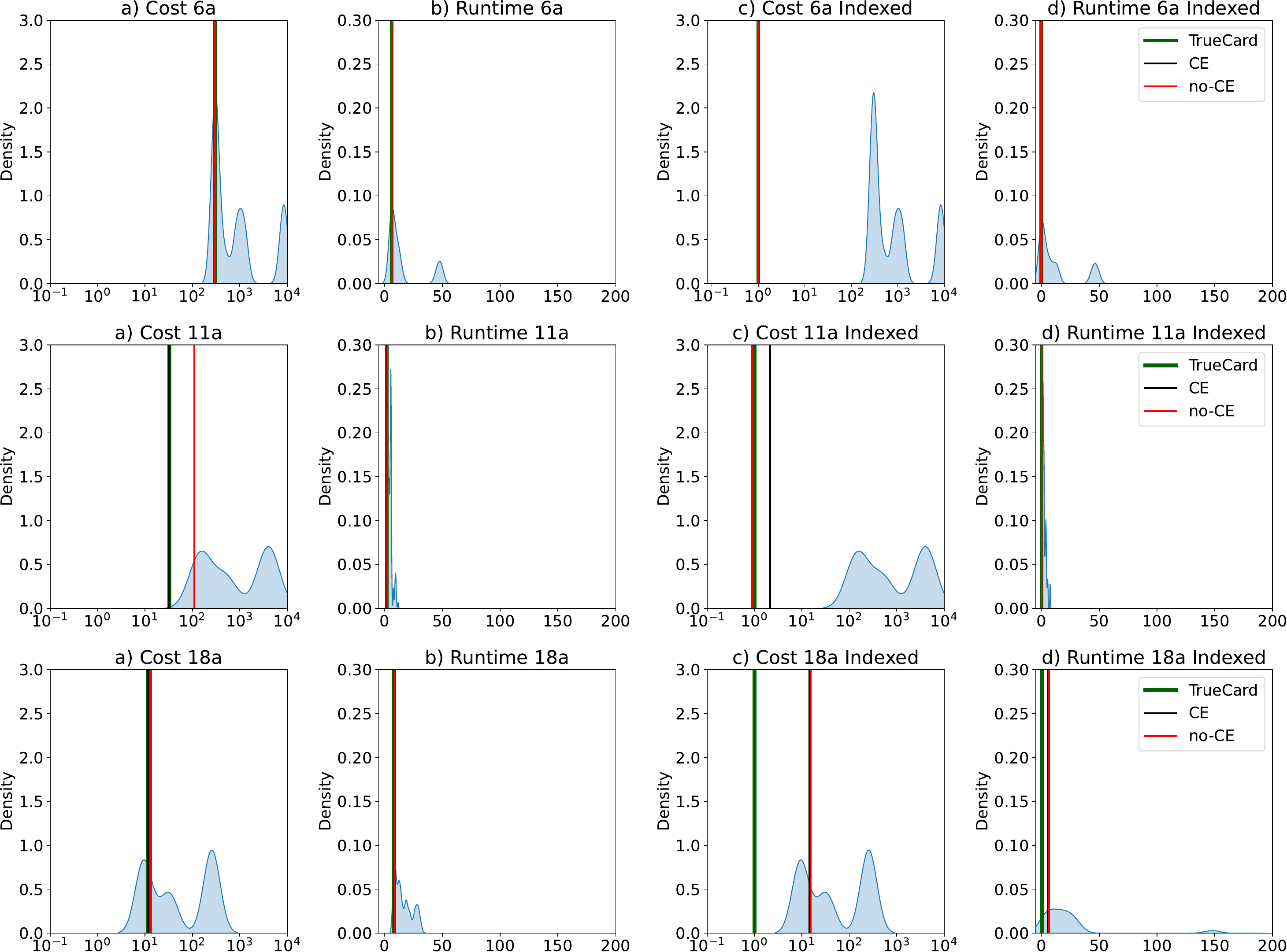}
	\caption{Cost and runtime distribution for query 6a, 11a and 18a. Cost relative to the TrueCard[log scale]. Runtime is in seconds.}
	\label{fig:quickpick}
\end{figure}

In our study, we generated 10,000 random plans for queries 6a, 11a, and 18a from the Join Order Benchmark (JOB)\cite{Leis:JOB:vldb-2018,Leis:QOREALLY:pvldb-2015} using the QuickPick algorithm\cite{Waas:JoinOrderSelection}. These plans are analyzed and visualized in Figure \ref{fig:quickpick}, where we display the distributions of plan costs and runtimes in both non-indexed and indexed settings. Each row in the figure depicts the cost and runtime distributions for a given query under both settings. The costs are normalized against the cost of TrueCard's indexed plan and are represented on a logarithmic scale. In each subplot, we highlight the costs or runtimes of plans derived from TrueCard, CE, and no-CE approaches with vertical lines in green, black, and red, respectively. The vertical lines coinciding indicate instances where these methods yield comparable costs or runtimes, as observed in sub-figure \ref{fig:quickpick} a), which focuses on the cost distribution for query 6a.

\noindent
\textbf{Non-Indexed - } 
Figure \ref{fig:quickpick}, column a, represents the cost distribution of all considered queries in a non-indexed setting. Similar cost of plans formed clusters. There are multiple such clusters of plans for the same query. For instance, query 6a demonstrates three distinct clusters of such plans. The cluster adjacent to the green line represents plans that are closest in cost to the optimal plan, whereas clusters to the right indicate progressively costlier plans. A closer inspection of the plans in the third cluster — the one farthest from the optimal cost — revealed a common pattern: many either commence with or incorporate a foreign key/foreign key join. This pattern persists across other queries as well. 

\begin{figure*}[htbp]
	\centering
	\subfigure[non-indexed setting.]{\label{fig:no-index-physical-operator}\includegraphics[width=.49\linewidth,height=40mm]{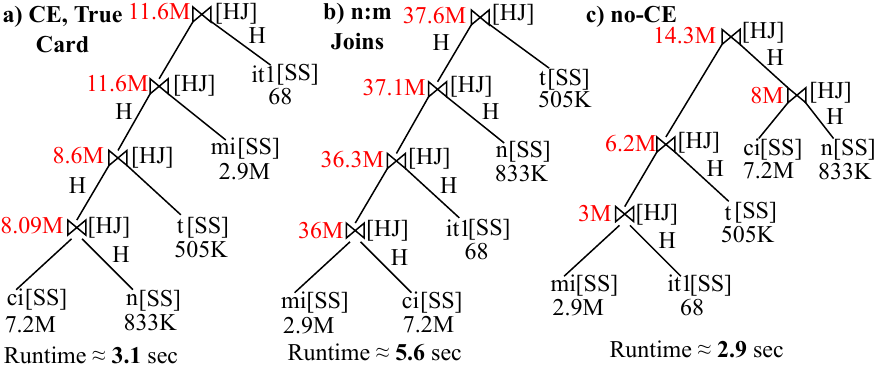}}
	\hfill
	\subfigure[indexed setting.]{\label{fig:indexed-physical-operator}\includegraphics[width=.49\linewidth,height=40mm]{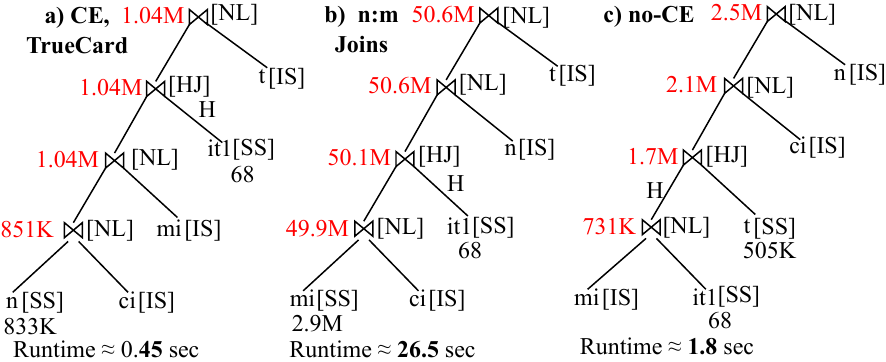}}
    \caption{Execution plans for query 18a.}
\end{figure*}

Figure \ref{fig:no-index-physical-operator} focuses on a specific subquery from query 18a to illustrate the effect of foreign key/foreign key joins on cost. It demonstrates that plans commencing with a join between foreign keys -- for example, \texttt{ci} and \texttt{mi} in this context -- tend to incur higher costs by generating large intermediate results early in the execution path. Contrastingly, in the other two scenarios analyzed within a non-indexed setting, such cost inflations do not occur.

Minimizing the occurrence of foreign key/foreign key joins could reduce the likelihood of generating plans that fall within this third, less cost-effective cluster for query 6a -- figure \ref{fig:quickpick}. Leis et al.\cite{Leis:JOB:vldb-2018,Leis:QOREALLY:pvldb-2015} argued that cardinality estimates are generally good for avoiding such joins. However, our research extends this discourse by illustrating how a no-CE approach can successfully avoid these joins even in the absence of cardinality estimates. 

Column b of Figure \ref{fig:quickpick}, which examines runtime, reflects a pattern analogous to that of cost, albeit with generally reduced distances between the clusters. This is particularly evident in the runtime distributions for queries 11a(b) and 18a(b), where the clusters are more tightly grouped compared to their cost counterparts, suggesting a lesser variance in the runtime efficiency of the plans.

\noindent
\textbf{Indexed - } In Figure \ref{fig:quickpick}, column c depicts the cost distribution of all considered queries in an indexed setting. Similar to non-indexed settings, similar plans cost form clusters for all queries. Notably, in the indexed context, these clusters tend to be positioned significantly to the right of the green line, which represents the TrueCard cost. This displacement suggests that the plans selected at random are generally associated with substantially higher costs compared to the TrueCard plan, a trend consistent across the queries studied. We assume that this phenomenon is largely attributable to the selection of physical operators, where errors in cardinality estimates can lead to sub-optimal choices, adversely affecting the cost of the plans. Conversely, in non-indexed settings, the choice of join physical operator was constrained to hash joins.

In the case of query 11a, the CE method demonstrates a higher cost in comparison to the TrueCard. With query 18a, both the CE and no-CE methods result in costs that exceed those associated with TrueCard. This underscores the pivotal role of precise cardinality estimates in settings where indexes are employed. Throughout our experiment, we fixed the join order in advance; however, the choice of physical operators was left to the discretion of the database management system, which in our experiments was PostgreSQL.

Column d of Figure \ref{fig:quickpick} presents the runtime analysis for the respective queries. The runtimes of similar plans tend to form clusters. Notably, multiple clusters are evident within the results for queries 6a and 18a. For query 11a, the clusters are more closely spaced, indicating a smaller variation in runtimes among the different plans.

\subsection{Physical Operators} \label{ssec:po-influence}

In Section \ref{ssec:no-ce-no-idx}, we examine the variance in cost when multiple alternatives for physical operators are available. The selection of appropriate physical operators is critical for generating efficient query plans. Various factors, such as cardinality estimates and system resource availability, influence this choice. For example, hash joins are favored when dealing with large cardinalities that can be accommodated in memory, whereas nested-loop joins may be more suitable for smaller cardinalities. However, imprecise cardinality estimates can lead to suboptimal selections of physical operators, thereby increasing query execution times. In this section, we investigate the impact of suboptimal physical operator choices on query runtime, even when the join order is optimal.

\noindent
\textbf{Non-Indexed - }
In the non-indexed configuration, we confine the optimizer to utilize hash joins exclusively, thereby eliminating the possibility of suboptimal plan selection due to inappropriate physical operator choices. As a result, join ordering becomes the sole performance bottleneck. Consequently, barring any optimizer errors—such as the early joining of two foreign key tables as illustrated in Figure \ref{fig:no-index-physical-operator}b—no significant performance degradation is observed, as evidenced by Figure \ref{fig:noidx_analysis}.

\begin{figure}[htbp]
	\centering
	\includegraphics[width = .6\textwidth]{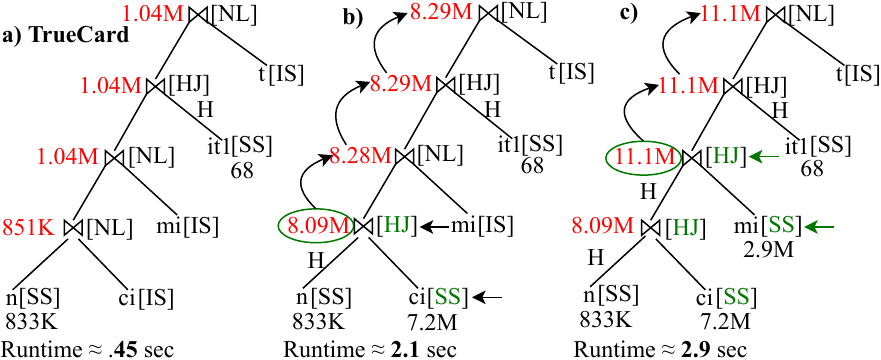}
	\caption{Impact of physical operators on optimal join order.}
	\label{fig:physical_operators_impact}
\end{figure}

\noindent
\textbf{Indexed - } The presence of indexes can substantially increase the range of physical operators available to the optimizer. When paired with the most suitable physical operators, TrueCard's plan cost reduces considerably compared to the non-indexed setting, as depicted in Figure \ref{fig:indexed-physical-operator}. The cost of the TrueCard plan dropped from 11.6 million to 1.04 million, and the runtime decreased from 3.1 to 0.45 seconds. To analyze the impact of physical operator choices on runtime, we made incremental alterations to scan and join operators in the TrueCard plan while keeping the join order intact. In Figure \ref{fig:physical_operators_impact}b, we change the scan and join operator for $ci$ from Index Scan(IS) to Sequential Scan(SS) and Nested Loop Join(NL) to Hash Join(HJ), respectively. This modification increased the plan cost by approximately eight times and the runtime by nearly 4.5 times. We repeat the process for $mi$ in Figure \ref{fig:physical_operators_impact}c. This change increased the plan cost by approximately eleven times and the runtime by roughly 6.5 times compared to TrueCard's. These results suggest that choosing suboptimal physical operators can significantly increase the overall runtime, even with an optimal join order. Therefore, selecting the most suitable physical operators is as important as determining the optimal join order for a query to achieve optimal performance.

\subsection{Parallel Processing and Runtime} \label{ssec:thread_runtime}

Contemporary query execution engines, including PostgreSQL version 9.6 and later, have embraced the capability for parallel processing. The degree of parallelism within a given query in PostgreSQL is governed by two parameters: \texttt{max\_parallel\_workers} and \texttt{max\_parallel\_workers\_per\_gather}. During the query optimization phase, PostgreSQL evaluates both parallel and sequential execution plans, in contrast to some other databases that may allocate threads dynamically at execution time. 

\begin{figure*}[htbp]
	\centering
	\includegraphics[width=\linewidth]{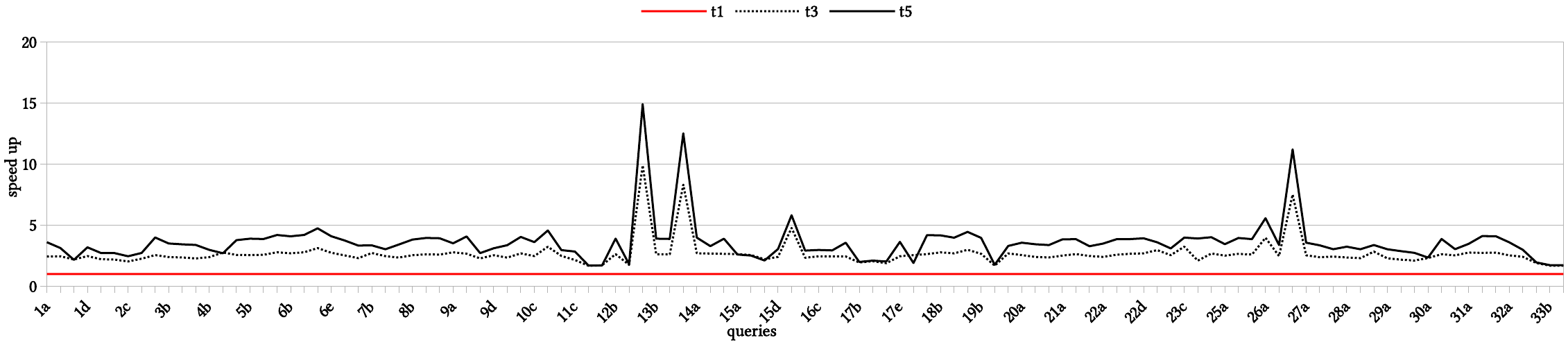}
	\caption{CE runtime ratio(speed up) for JOB queries between multi-threaded(t3, t5) and single-threaded(t1) - Non-indexed setting}
	\label{fig:thread_runtime_noidx}
\end{figure*}

\begin{figure*}[htbp]
	\centering
	\includegraphics[width=\linewidth]{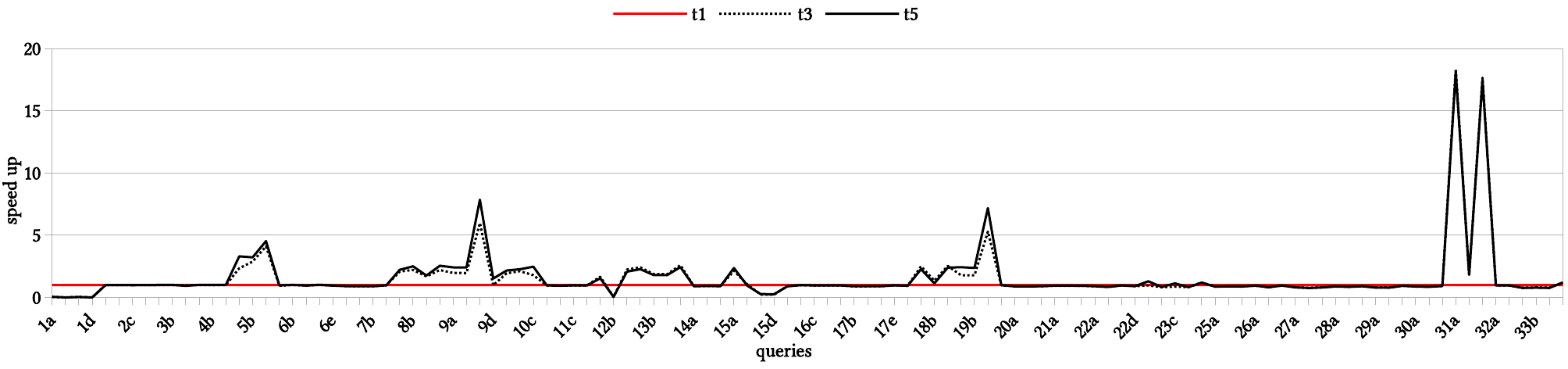}
	\caption{CE runtime ratio(speed up) for JOB queries between multi-threaded(t3, t5) and single-threaded(t1) - indexed setting}
	\label{fig:thread_runtime_idx}
\end{figure*}

Figure \ref{fig:motivation} illustrates the variability in execution times for the Join Order Benchmark (JOB) workload under the CE method as thread count increases, a scenario examined in both indexed and non-indexed configurations. To delve deeper into these observations, we analyze the proportional disparity in runtime between multi-threaded and single-threaded (t1) executions as depicted in figure \ref{fig:thread_runtime_noidx}, \ref{fig:thread_runtime_idx}. We further investigate to identify which queries benefit from parallel execution and which do not for both the CE and No-CE methods. Our experiments span a range from single-threaded (t1) to five-threaded (t5) execution, with the empirical finding that extending beyond five threads yields negligible performance improvements in PostgreSQL.

\noindent
\textbf{Non-Indexed - } In Figure \ref{fig:thread_runtime_noidx}, an analysis in the non-indexed environment reveals that the runtime speed up for a three-threaded (t3) configuration exhibits a range from 1.68 to 9.90 throughout the JOB workload. Advancing to a five-threaded (t5) scenario, the amplification in runtime efficiency ascends to a maximum of 14.92. The speed-up is computed utilizing the formula (t1 runtime)/(t3 runtime) to denote the speed-up for t3. Transitioning from a single-threaded(t1) to a five-threaded(t5) environment results in a runtime reduction of 627.04 seconds for the CE approach and approximately 750.3 seconds for the No-CE approach as detailed in Table \ref{tab:improvement}. This increase in efficacy primarily stems from the adept parallelization of physical operations, particularly Sequential Scans and Hash Joins, within non-indexed scenarios. These findings underscore that in the absence of indexes, an elevation in thread count exerts a positive impact on query runtimes, though the magnitude of acceleration varies across individual queries.

\begin{table}[htbp]
    \centering
    \begin{tabular}{@{}llrrrr@{}}
    \toprule
	\textbf{method} & \textbf{setting} & \textbf{<0 \%} & \textbf{0.1 - 39 \%} & \textbf{40-94 \%} & \textbf{improvement percentage}\\ \midrule
	\multirow{4}{*}{CE} & \multirow{2}{*}{non-indexed} & 0 & 0 & 100 & \% of queries \\
    	& & 0 & 0 & 627.04 & runtime decrease in seconds \\ \cmidrule{2-6}
		& \multirow{2}{*}{indexed} & 62 & 13.20 & 23.90 & \% of queries \\
    	& & -12.50 & 1.96 & 48.78 & runtime decrease in seconds \\ \midrule
	\multirow{4}{*}{No-CE} & \multirow{2}{*}{non-indexed} & 0 & 1.76 & 98.23 & \% of queries \\
    	& & 0 & 1.48 & 748.82 & runtime decrease in seconds \\ \cmidrule{2-6}
		& \multirow{2}{*}{indexed} & 39.80 & 6.10 & 54 & \% of queries \\
    	& & -2.31 & 0.81 & 490.3 & runtime decrease in seconds \\ \bottomrule
    \end{tabular}
    \caption{Variation in query execution time (Seconds) as a function of thread count increment from 1 to 5, with the proportion of queries per category represented as a percentage of the total 113 JOB queries.}
    \label{tab:improvement}
\end{table}

\noindent
\textbf{Indexed - } In an indexed configuration, the comparative analysis between a five-threaded (t5) and a single-threaded (t1) execution yields a diverse spectrum of results. Specifically, under the CE approach, 62\% of queries experience a degradation in performance, 13.20\% exhibit negligible changes, and a noteworthy enhancement in runtime—exceeding 40\%—is observed in 23.90\% of queries. The performance variance is quantified using the following formula:

\[
\text{Performance Change} = \left( \frac{\text{Runtime}_{t1} - \text{Runtime}_{t5}}{\text{Runtime}_{t1}} \right) \times 100\%
\]

Similarly, while using the No-CE approach, a performance decrement is observed in 39.80\% of queries, while 6.10\% remain essentially stable, and an improvement exceeding 40\% is discernible in 54\% of queries. Both the CE and No-CE approaches indicate that a significant number of queries fail to exhibit performance improvements, a pattern that is documented in Table \ref{tab:improvement}. To elucidate the underlying dynamics at play, we turn our attention to the operator statistics presented in Table \ref{tab:op_indx}.

Table \ref{tab:op_indx} categorizes the JOB workload into two distinct groups based on performance trajectory: (a) <0, where performance wanes, and (b) >0, where performance is improved relative to a single-threaded baseline (t0). Within the first category, the adoption of physical operators remains invariant from t1 through t5. Index Scans are the preferable choice for scan operation in most cases, followed by Sequential Scans. Concurrently, join operations are primarily dominated via Nested Loop Joins for both thread configurations t1 and t5. The invocation of Hash Join operations is relatively infrequent, registering at 15 instances for t1 and 19 for t5. We observe a similar trend for the No-CE approach. 

\begin{table}[htbp]
    \centering
    \begin{tabular}{@{}lllrrrrr@{}}
    \toprule
	\textbf{method} & \textbf{improvement} & \textbf{threads} & \textbf{Index Scans} & \textbf{Heap Scan} & \textbf{Seq Scan} & \textbf{Nested Loop} & \textbf{Hash Join} \\ \midrule
	\multirow{4}{*}{CE} & \multirow{2}{*}{a) <0(71 queries)} & 1(t1) & 450 & 66 & 209 & 573 & 15 \\
    	 & & 5(t5) & 441 & 61 & 211 & 569 & 19 \\ \cmidrule{2-8}
		 & \multirow{2}{*}{b) >0(42 queries)} & 1(t1) & 212 & 22 & 106 & 231 & 45 \\
    	 & & 5(t5) & 218 & 11 & 107 & 214 & 62  \\ \midrule
	\multirow{4}{*}{No-CE} & \multirow{2}{*}{a) <0(45 queries)} & 1(t1) & 294 & 29 & 116 & 344 & 21 \\
    	 & & 5(t5) & 290 & 24 & 120 & 339 & 26 \\ \cmidrule{2-8}
		& \multirow{2}{*}{b) >0(68 queries)} & 1(t1) & 304 & 14 & 263 & 320 & 179 \\
    	 & & 5(t5) & 302 & 2 & 265 & 300 & 199  \\ \bottomrule
    \end{tabular}
    \caption{Impact of thread count on operator statistics for queries with improvement and decline in execution times.}
    \label{tab:op_indx}
\end{table}

Category (b) reflects a pattern akin to category (a), with Index Scans and Nested Loop Joins being the prevalent methods for scan and join operations, respectively. However, category (b) distinguishes itself by a greater incidence of Hash Joins—constituting 22\% of joins in CE(t5) and 40\% in No-CE(t5) — figures that surpass those observed in category (a), where Hash Joins account for a mere 3\% in CE and 7\% in No-CE. Within the CE(t5) setting, a notable 23.9\% of JOB queries exhibit performance gains ranging from 40\% to 94\% over the single-threaded baseline (t0), culminating in a runtime reduction of 48.78 seconds. In contrast, the No-CE approach witnesses 54\% of queries outperforming t0 by 40\% to 94\%, resulting in a substantial runtime decrease of 491 seconds.

Despite the inherent capability of all operators listed in Table \ref{tab:op_indx} to capitalize on parallel processing, category (a) does not demonstrate any performance gains, whereas category (b) shows a moderate improvement under CE and a more pronounced enhancement in No-CE. We assume that the observed performance improvements are attributable to the heightened utilization of hash join operations, which appear to be more amenable to parallel execution in this case.

\subsection{In-memory experiments} \label{ssec:in-memory}
To ensure a comprehensive evaluation, our study examines the performance of modern main-memory databases, specifically DuckDB, HEAVY.AI, and MonetDB, a columnar database. While MonetDB does not offer guaranteed join ordering, our analysis aligns with prior work \cite{Hertzschuch:simplicity:cidr-2021} and incorporates its findings. We present the execution times for non-indexed configurations in Table \ref{tbl:in-memory-runtimes}. We omitted the indexed settings from our results due to the absence of comprehensive index support in HEAVY.AI,  coupled with minimal or no performance gains observed in DuckDB and MonetDB when utilizing indexes. Our evaluation considers both the default runtime for each database system, indicated as 'default' in Table \ref{tbl:in-memory-runtimes}, as well as execution plans chosen by the no-CE algorithm.

DuckDB utilizes a basic query optimizer, considering join types and column distinct value counts, differing from no-CE, which does not rely on statistics. In our experiments, DuckDB completed the JOB workload in 30.38 seconds, while no-CE's duration was 37.89 seconds in DuckDB. HEAVY.AI struggled with the same workload, timing out on 37 out of 113 queries based on a 120-second threshold. Accounting for these timeouts by assigning a 120-second penalty for each, the cumulative time for the JOB workload amounted to 4744 seconds. This performance shortfall was anticipated, given that HEAVY.AI does not optimize table arrangements for joins but rather executes queries in their given order. When we run no-CE queries in HEAVY.AI, the JOB workload completion time was 282.50 seconds. MonetDB, on the other hand, employs a more advanced, cost-based optimizer that utilizes both statistics and heuristics, thereby distinguishing itself from both DuckDB and HEAVY.AI. In our tests, MonetDB processed the JOB queries in 114.46 seconds, but when augmented with no-CE method, the time was further reduced to 88.79 seconds.

In the context of hardware-accelerated databases, we find that the performance of the no-CE method is on par with the default optimization strategies in JOB. This observation is likely attributable to the databases' minimal reliance on cardinality estimates during the optimization process. 

\begin{table}[htbp]
	\centering
	
	\begin{tabular}{lrr}
	\toprule
			 & \textbf{default}           & \textbf{no-CE} \\ \midrule
	DuckDB   & 30.38       		 		  & 37.87             		   \\ 
	HEAVY.AI & \textgreater 4744 		  & 282.50            		   \\ 
	MonetDB  & 114.46            		  & 88.79             		   \\ \bottomrule
	\end{tabular}
	\caption{Execution times (seconds) in DuckDB, HEAVY.AI and MonetDB.}
	\label{tbl:in-memory-runtimes}
\end{table}

\subsection{Insights}\label{ssec:insights}
Throughout this paper, we have conducted a comprehensive analysis of query optimization with and without cardinality estimates, leading to several observations. Some of these observations build upon existing knowledge from published database systems literature, while others are new. In the following, we list these findings according to their respective Sections:

\textbf{Section \ref{ssec:analysis-ceVSnoce},\ref{ssec:no-ce-no-idx}, \ref{ssec:in-memory} -} The performance disparity between TrueCard, CE, and no-CE approaches is notably marginalin a non-indexed environment in the Join Order Benchmark. The CE and no-CE methods frequently perform on par with TrueCard, with only a few deviations. To understand this outcome, we randomly select 10,000 plans using the QuickPick algorithm for queries 6a, 11a, and 18a. The plan cost distribution forms multiple clusters, where each cluster corresponds to plans with similar costs. A closer inspection of those plans that incur costs significantly above the TrueCard frequently initiates with or include a join between foreign keys. Leis et al.\cite{Leis:JOB:vldb-2018,Leis:QOREALLY:pvldb-2015} argued that cardinality estimates are generally good for avoiding such joins. However, our research extends this discourse by illustrating how a no-CE approach can successfully avoid these joins even in the absence of cardinality estimates. 

\textbf{Section \ref{ssec:opt-time} -} The planning time of CE grows exponentially with the number of joins in both non-indexed and indexed configurations. As the number of joins increases, the search space expands, resulting in a longer search time to identify the optimal plan. In contrast, no-CE utilizes heuristics to find the optimal plan, leading to a nearly consistent planning time across different queries. This observation suggests that planning time becomes a bottleneck for CE when the number of joins increases. Therefore, it is crucial to fine-tune the threshold for the transition between dynamic programming and genetic algorithms.

\textbf{Section \ref{ssec:po-influence} -} Numerous research works\cite{Marcus:DRLJOE:aiDM-2018,Leis:JOB:vldb-2018,Leis:QOREALLY:pvldb-2015,Steinbrunn:HRO:jvldb-1997,Waas:JoinOrderSelection,Hertzschuch:simplicity:cidr-2021} have focused on identifying the optimal join order, yet studies on optimal physical operator selection remain limited \cite{Hertzschuch:physical_operators:vldb_2022}. In Section \ref{ssec:po-influence},  we present an example illustrating the impact of suboptimal physical operators on optimal join order. These highlight the importance of selecting optimal physical operators in conjunction with optimal join order to achieve the best possible query performance.

\textbf{Section \ref{ssec:thread_runtime} -} In database research, studies involving CE methods often rely on single-threaded configurations\cite{Leis:JOB:vldb-2018, Han:CEB:2021, Han:CEB-github:2021}, leaving the impact of varying thread counts largely unexplored. In Section \ref{ssec:thread_runtime}, using the Join Order Benchmark, we analyze the effect of the number of threads on query runtime in both indexed and non-indexed settings. We observe that as the number of threads increases, runtime improves, and the performance discrepancy between TruCard and CE, no-CE tends to diminish.

\section{Related work}\label{sec:related-work}   

Query optimizers can be broadly categorized into two distinct types based on their input characteristics: cardinality estimation-based and heuristic-based optimizers.

\noindent
\textbf{Cardinality estimation based optimizer.}
Cardinality estimation is crucial for determining optimal query execution plans. Leis et al.\cite{Leis:JOB:vldb-2018,Leis:QOREALLY:pvldb-2015} explored the impact of cardinality estimates on query optimization. Estimates often suffer from inaccuracies due to oversimplified assumptions like uniformity, independence, and inclusion. PostgreSQL~\cite{postgres} uses histograms for data representation, relying on formulas based on these assumptions. Although histograms work well for single attribute estimations, they struggle with join-crossing correlations. Cai et al.\cite{Cai:PCETUB:sigmod-2019} introduced Pessimistic, employing count-min sketches to capture foreign key join-crossing correlations, but the sketch-building process introduces significant overhead as join numbers increase. Hertzschuch et al.\cite{Hertzschuch:simplicity:cidr-2021} maintained pessimistic cardinality estimation properties while substituting sketches with an upper bound formula leveraging statistics already available to PostgreSQL. Izenov et al.\cite{Izenov:compass:sigmod-2021} used Fast-AGMS sketches to capture join-crossing correlations, reducing overhead during sketch-building compared to Pessimistic. 

In recent years, machine learning has gained significant interest in the field of query optimization. Numerous studies have focused on cardinality estimates~\cite{Kipf:LCECJDL:cidr-2019,Marcus:NEO:pvldb-2019,Woltmann:CEL:aiDM-2019,Liu:CEUNN:cascon-2015,Ortiz:EADLCE:arxiv-2019,Malik:BBAQCE:cidr-2007}, while relatively fewer approaches have concentrated on join-ordering~\cite{Krishnan:LOJQDRL:arxiv-2018,Marcus:DRLJOE:aiDM-2018} or end-to-end query optimization~\cite{Marcus:NEO:pvldb-2019}. Though machine learning techniques demonstrate accuracy for single-table cardinality estimates~\cite{wang2021ready}, their performance tends to decrease in multi-table queries, as shown by Han et al.~\cite{han2021cardinality}. These machine-learning models often necessitate extended training time, which can limit their practical applicability in certain scenarios.

\noindent
\textbf{Heuristic based optimizer.}
Heuristic-based query optimizers employ predefined heuristics to identify the optimal plan for query execution. Several heuristic-based systems\cite{ingres,Starburst,orcl-qop,Volcano,EXODUS,Praire,venusdb,calcite,presto} have developed their own rule languages and execution environments to avoid compatibility issues. Held et al. \cite{ingres} introduced Ingres, the first rule-based system, where the original query is divided into single-valued sub-queries and executed separately using a greedy approach. While effective for simple queries, this method struggles with complex queries. In contrast, Pirahesh et al. \cite{Starburst} developed Starburst, a Query Graph Model (QGM) based system that represents a SQL query as a graph. Query rewriting rules transform one QGM into an equivalent QGM, and during the plan optimization phase, each equivalent QGM is assigned an estimated cost, with the lowest cost QGM selected for query execution. Integrating query graphs with join types and functional dependencies may help to find efficient plans for execution in a main-memory setup\cite{Datta:Simpli2:arXiv-2021}. Graefe et al. \cite{EXODUS} created EXODUS, where a query is represented as an algebraic tree and employs rule-based reordering and plan optimization techniques similar to Starburst. However, the simplistic search strategy and cost function used by Starburst and EXODUS introduce limitations for complex queries. To address these limitations, Graefe et al. \cite{Volcano} presented Volcano, which utilizes directed dynamic search rather than rules for enumeration.

\section{CONCLUSIONS}\label{sec:conclusions}

This paper presents an in-depth analysis of query optimization with and without cardinality estimates, examining the interplay between join orders, physical operators, and parallelism. Our findings demonstrate that the performance gap between TrueCard and CE, no-CE, is minimal in non-indexed settings and main-memory databases. Furthermore, we highlight the importance of fine-tuning the threshold for transitioning between dynamic programming and genetic algorithms as CE's planning time increases substantially with the number of joins. Our study also highlights the crucial role of optimal physical operator selection in conjunction with join order for maximizing query performance. Our investigation extends to assess the effects of parallel processing on query performance, examining its implications across both indexed and non-indexed settings. Overall, our findings contribute to a deeper understanding of query optimization in the presence and absence of estimates.

\paragraph*{Acknowledgments.}
This work is supported by NSF award number 2008815.

\bibliographystyle{plain}
\bibliography{biblio}  

\end{document}